\magnification=\magstep1
\vsize=8.5truein
\hsize=6.3truein
\baselineskip=20truept
\parskip=4truept
\vskip 18pt
\def\today{\ifcase\month\or January\or February\or
March\or April\or May\or June\or July\or
August\or September\or October\or November\or
December\fi
\space\number\day, \number\year}
 
\centerline{\bf Extended Capillary Waves and the Negative Rigidity Constant }
\centerline{\bf in the d=2 SOS Model }
\medskip
\centerline{by }
\centerline{J. Stecki }
\vskip 20pt
\centerline{Department III, Institute of Physical Chemistry,}
\centerline{ Polish Academy of Sciences, }
\centerline{ ul. Kasprzaka 44/52, 01-224 Warszawa, Poland}
\bigskip
\centerline{\today}
\vskip 60pt
\centerline{\bf{Abstract}}

The solid-on-solid (SOS) model of an interface separating two phases 
is exactly soluble in two dimensions (d=2) when the interface becomes
a one-dimensional string. The exact solution in terms of the transfer matrix 
is recalled and the density-density correlation function $H(z_1,z_2;\Delta x)$
together with its projections, is computed.  It is demonstrated
that the shape fluctuations follow the (extended)  capillary-wave theory expression
$S(q)=kT/(D+\gamma q^2 +\kappa q^4) $  for sufficiently small 
wave vectors $q$. We find  $\kappa$   {\it negative}, $\kappa <0$ .
At $q=2\pi$ there is a strong nearest-neighbor peak.  
Both these results  
confirm the earlier findings as established in simulations in 
d=3 and in continuous space, but now in an exactly soluble lattice model.
\vfill\eject 
{\bf I.Introduction}

The shape fluctuations of interfaces are described by the capillary-wave 
theory[1-10]. Its principal result is that the Fourier spectrum of the 
fluctuations follows for small wave-vectors the 
$1/q^2$ divergence. The divergence  may be damped by an external field 
or by finite size, as in the 
form $1/(D+\gamma q^2)$, valid in the limit $q\rightarrow 0$.
Both $D$ and the interfacial tension $\gamma$ are positive. The scattering 
experiments scan an interval of wave vectors up to the nearest-neighbor 
peak and beyond and  many authors have searched for an extension of the 
simple capillary-wave form. An obvious generalization appeared to be
$$ S(q) =kT/(D+\gamma q^2 +\kappa q^4). \eqno (1.1) $$
Sum rules require convergence of integrals over $q$ and $\kappa>0 $ 
seems natural. Additional input was provided by the fluctuations of 
other $(d-1) - $dimensional objects in $d-$dimensional systems, such as 
membranes and bilayers in liquid solvents. According to both  
theory and experiment,
  the fluctuations of those autonomous objects are ruled not 
by $\gamma$ but by rigidity coefficient coupled to the {\it curvature}.
It seemed only natural to "improve" the theory of fluctuations of 
ordinary non-autonomous interfaces by incorporating a contribution of a 
hypothetical rigidity coefficient coupled to  curvature. The latter 
contribution ought to be 
$$ kT/\kappa q^4 \eqno(1.2)$$ 
with $\kappa$ necessarily positive. The positivity condition is necessary 
for the free energy of the deformed state to be larger than the free 
energy of the undeformed state. In this way one arrives at the expression 
(1.1)  with the expectation of all three constants being positive.

   The first signs of inconsistency were signalled in a series 
of papers[8] where it was demonstrated that 
(for the long range attractive dispersion force, but not only) 
expression (1.1) is inadmissible and the last term 
must be replaced by $\kappa q^4 f(q)$ where the function $f$  contains 
nonalytical terms with logarithms.   

  In a series of simulations 
by Molecular Dynamics of liquid-vapor surfaces,  
it was found independently[6] that all
results pointed to a negative $\kappa$. A negative value
disproved the interpretation of $\kappa$ as a rigidity coefficient. 
The unexpected results were recently confirmed by simulations 
of capillary waves in other liquid systems[9]. In either case[6,9] the
intermolecular attractions between particles, were technically of short 
range because of the cutoffs. 

   It may  be argued that these two independent simulation results[6,9] 
are also confirmed by scattering experiments; a negative $\kappa$ means that 
the denominator gets smaller 
than it would have been without any $\kappa$,  
the scattering intensity gets 
bigger - and this was observed for liquid-vapor surfaces[10]. 
Such an interpretation, however, is not generally accepted and other,
more complicated, explanations were sought[5,10].

 It would be interesting to confirm or negate the finding 
of a negative $\kappa$  by a method other than simulation, for example
in a system or model which is  soluble exactly. 

\bigskip
{\bf II. The SOS model. }

In the Solid-on-Solid (SOS) model in two dimensions d=2 [11-17],
a microscopic configuration of the interface is given by 
the collection of heights
$$ \{h_i\}, i=1,L ;~~ 1\le h_i \le M \eqno(2.1)$$
as illustrated in Fig.1. 
The strip is  $M\times L$ and below we  consider the 
limit $L \gg M $. The other limit, $M \gg L $, has also been considered[14,17].
The "vertical" variable is $h$ or $z$ at the horizontal position 
$x=a_0 j$, $0<j\le L$ and we take $a_0=1$. The "height" is limited 
to  $0\le h\le M$.
The standard $d=2$ SOS model[11,17] assumes the energy of a  
given configuration as
$$ E=\epsilon \sum \vert h_{i+1} - h_i \vert ^n~~~~\epsilon >0~ \eqno(2.2)$$
with $n=1$. A Gaussian model ($n=2$) can also be considered.
Introducing the $M\times M$ transfer matrix
$$ T(h,h')=Q^{\vert h-h' \vert} \eqno(2.3)$$
with $Q\equiv \exp[-\epsilon /kT] $, we find that the
partition function and correlation functions can be expressed
in terms of ordinary scalar products of matrices $T$. If the 
transfer matrix is diagonalized - as it can be - all solutions are
available, also as analytical expressions[11][14][17][15]. There  is a 
minor and unimportant approximation involved in  the analytical 
expressions for the eigenvectors and eigenvalues of the inverse of $T$[15]
(see Appendix).

We have calculated the density-density correlation function
$\tilde H(z_1,z_2;q)$ and other quantities such as 
the density profile $\rho(z)$ and
the probability functions: one-point $p(h)$ and two-point 
$P(h_1,h_2;\Delta x)$ and $\tilde P(h_1,h_2;q)$.
The following relations take place
$$ p(h)=\rho(h+1)-\rho(h);~~ p(h)= Z(h,1)^2 \eqno(2.4)$$
where $Z_1(h)=Z(h,1)$ is the dominant eigenvector corresponding to the 
largest eigenvalue $\lambda_1$. Moreover 
$$  H = \sum_{h_1\ge z_1}\sum_{h_2\ge z_2} P(h_1,h_2)-p(h_1)p(h_2)\eqno(2.5)$$
at each value of $\Delta x \equiv \vert x_2 - x_1 \vert$. 
Calculation of the Fourier Transform gives
$$ \tilde H(z_1,z_2; q) = \sum_{h_1\ge z_1}\sum_{h_2\ge z_2}
\sum_{j\ge 2} f_j(q) c_j(h_1)c_j(h_2), \eqno(2.6)$$
where
$$ c_j(h)= Z_j(h)Z_1(h). \eqno(2.7)$$
In
$$ f_j(q)=(1-r^2)/(1+r^2 -2r \cos (q) ), \eqno(2.8)$$
$r\equiv r_j = \lambda_j/\lambda_1 < 1 $ and we never need 
$j=1$.  Periodic boundary conditions in the $x-$direction 
were assumed, $h(x=1) = h(x=L+1)$; next
the limit $L\rightarrow \infty$ has been taken at constant $M$.
 It is useful for the 
discussion of the $q-$dependence to introduce projections 
which result in  functions of one variable, $q$. One such function is 
$\bar H(q)$ which projects, at each $q$, the 
matrix $\tilde H$ onto $p(h)$.
A similar projection was used in d=3[2,6].
$$ \bar H = \sum_{h_1}\sum_{h_2} p(h_1) H(h_1,h_2) p(h_2). \eqno(2.9) $$
 Since $p(h)=Z_1(h)^2$, by using (2.5) we can change the order of summations
to make a computationally convenient expression.

Another projection is the height-height correlation function 
$<h(x_1)h(x_2)>$ whose Fourier transform is the well known structure
factor (of the interface)
$$ S(q) = <h_q h_{-q} >.   \eqno(2.10)$$
We have 
$$<h(x_1)h(x_2)>= \sum_h \sum_{h'} h h' [P(h,h';\Delta x) -p(h)p(h')] \eqno(2.11)$$
and,  constructing the Fourier sums,
$$ S(q)=\sum_{h}\sum_{h'}\sum_{j\ge 2}  f_j(q) c_j(h)c_j(h')h h' , \eqno(2.12)$$
where the largest eigenvalue is excluded from  the sum over the 
eigenvalue index $j$.
Alternatively  the definition (2.6) of $c_j$ can be invoked and the 
order of summation changed to give
$$ S(q) =\sum_{j\ge 2}  f_j(q) d_j(q)^2   \eqno(2.13)$$
where
$$  d_j(q)\equiv \sum_h Z_j(h)  h Z_1(h).  \eqno(2.14)$$
To summarize, we have so far the full density-density correation function
$\tilde H(z_1,z_2,q)$, the corresponding  $z-$derivative $P(h_1,h_2,q)$
and the projections $\bar H(q)$ and $S(q)$. Moreover, for a given value of 
$q$, the density-density correlation function $\tilde H$ is a $M-1 \times M-1$ 
positive definite matrix; it can be diagonalized and all its $M-1$ 
eigenvalues are functions of $q$; the largest eigenvalue can be
chosen and discussed as function of $q$.

 To make practical use of these expressions, the $T-$matrix is 
diagonalized as function of $Q$ and this can be done analytically for
small values of $M$, or one can take the route through the inverse of $T$,
which is easier because the inverse is tridiagonal[11]. The final expressions are
given in the appendix[15].
\bigskip
{\bf III. Results for the wave-vector dependence of correlations.}

In order to reveal and illustrate the details of the capillary waves in the SOS 
model recalled and defined in the previous section, we  
numerically compute  the analytical expressions given in Section II 
and in the Appendix. In the Figures we show a selection of data, all for 
$Q=0.5$. 

 The numerical computation of $\tilde H$, of the projection 
 $\bar H$, and of the interfacial (string) structure factor $S(q)$, 
reveals three regions 
of the variable $q$. These regions are clearly visible in 
 Fig.2. At the very lowest values of $q$ a plateau (in logarithmic scale)
practically equal to the limit for $q=0^+$ extends up to a crossover
into the second and main region, that of the capillary 
waves. All the projections show a typical capillary-wave dependence of 
the form of (1.2), $1/const.q^2 $. 
This region extends up to $q^2\sim 1-5 $ where the amplitudes of the capillary 
waves have
practically died out. Then in the third region ($q^2 > 5$) 
the nearest neighbour peak appears near $q\sim 2\pi$ i.e. in Fig.2
near $q^2 \sim 39$~ .  

The plateau and the finite value at
$q=0^+$ are due to the finiteness of the width  $M$ in 
the $z-$direction perpendicular to the interface. The excursions of the
interface from the average position in the middle at $\sim M/2$ are
 damped like waves in a channel, of width $M$ and length $L$. 
This is captured by the smallest values of $q$.

The main capillary-wave region dominates in the middle region
and is the main physical effect.

    The wave-vector $q$ is measured in units of inverse lattice 
 constant $a_0$, taken as unity here; hence the nearest neighbour peak 
 expected at $2\pi/a_0$ occurs at $x=q^2 \sim 4\pi^2$. Its appearance 
 was never considered in earlier work. 
 
 This simple fact is, however, of importance because
 the presence of the peak requires {\it a raise} in the structure 
 factor at  $q<2\pi$, and this gradual raise takes place already 
at values of $q$ significantly lower than $2\pi$. 

 It may be found surprising that the raise begins so early.

 The plots of $\bar H$ for different sizes (widths) are shown in 
 Fig.3. 
 The  plot of $S(q)$ in Fig.4 shows a curve common to 
 all sizes in the central (capillary-wave) portion of the plot 
 and the raise of $S(q)$ preceding 
 the n.n. peak. Each of the $\bar H(q)$ and $S(q)$ for each size 
 $M$    
 were fitted to an empirical version of (1.1), i.e. to
 $$ S(x)= 1/(d + g x + K_4 x^2)~~~ x\equiv q^2 \eqno(3.1) $$
 where the constants $d,g,K_4$ or $g,K_4$ were estimated from the least-square
 fits. 
 In all cases but one the least-square fit produced a negative 
 $K_4$. The same formula was used for fitting  the 
 largest eigenvalue of $\tilde H(z_1,z_2;q)$ and invariably negative 
 $K_4$ was obtained.
 The raise in S(q) can be mistakenly attributed to other reasons,
 if the existence of the faraway nearest-neighbour peak is neglected.

 \bigskip
 {\bf IV.Discussion}
 
We have found in this exactly soluble lattice model in d=2 dimensions 
the same feature discovered earlier in two[6,9] independent simulations:
a  raise in the structure factor over and above the capillary-wave 
contribution - which results in a {\it negative} value of the 
coefficient $\kappa$ in (1.1), to which the 
empirical coefficient $K_4$ is proportional . 
Therefore it cannot be 
interpreted as the rigidity of the interface.

There exist systems with capillary waves and positive  $\kappa$ : 
as mentioned in the introduction these 
are bilayers (and membranes) in d=3 immersed in a liquid solvent. 
For each such  autonomous object there  exists a surface area which
minimizes the free energy.  
No such minimum exists in non-autonomous interfaces nor in the Ising or
SOS model.

In general we must conclude that there exist several kinds of interfaces 
ruled by different laws. To strenghten this point one can invoke 
the existence of Langmuir monolayers with their numerous 
surface phases.

  We may associate with the bilayers and membranes the term 
  "autonomous" and with ordinary liquid-vapor and liquid-liquid
  interfaces the term "non-autonomous". 
\bigskip
{\bf Acknowledgements.}

  The author is indebted to Professors Alina Ciach, Ben Widom, and 
  R.Evans, FRS,  for a discussion. 
\vfill\eject

{\bf Appendix A.}

The $T$-matrix given by (2.3) produces on inversion a tridiagonal 
matrix $\tau$ with  diagonal elements given by 
$$ d =(1+Q^2)/(1-Q^2)    \eqno(A.1) $$
and all offdiagonal elements $\tau_{n,n+1},\tau_{n,n-1}$ given by
$$ o = -Q/(1-Q^2)    \eqno(A.2) $$
Strictly speaking not all diagonal elements are equal to $d$; 
the first and the last element are
$$ \tau_{11}=\tau_{MM}=d_0=1/(1-Q^2)  \eqno(A.3) $$
For large $M$, safely $d$ can be substituted for $d_0$ and the 
resulting matrix $\tau'$  has the following eigenvalues[15]
$$ \Lambda_j =(1+Q^2 -2Q\cos (b_0 j) )/(1-Q^2)  \eqno(A.4) $$
where $b_0= \pi/(N+1)$ and $j = 1,\cdots ,M$.
Hence j=1 corresponds to the smallest eigenvalue, i.e. to the 
largest eigenvalue of the original matrix $T$, equal to $1/\Lambda_1$.
The eigenvectors 
are 
$$ Z_j(h) = (1/{\cal N})\cos(b_0 j \nu)       \eqno(A.5) $$
for j odd , and 
$$ Z_j(h) = (1/{\cal N})\sin(b_0 j \nu)     \eqno(A.6) $$
for j even. These eigenvectors are orthonormal and complete.
The dimensions of the matrices are $M \times M$ and $M$ is best taken odd.
The norm is $ {\cal N}^2 = (M+1)/2$ and $\nu= h - 1 -(M+1)/2$~ so that 
$\nu = -(M+1)/2,...,0,...,+(M+1)/2$.  

The eigenvectors are common to the diagonalized matrix and its
inverse whereas the eigenvalues are the inverses. Because $T$ is a
symmetric positive definite matrix, it admits a representation
$$ T_{h_1,h_2}= \sum_j Z_j(h_1)(1/\Lambda_j) Z_j(h_2)  \eqno(A.7) $$
and this representation for $T$ was used to obtain expressions in
Section II, remembering that (A.4) refers to the inverse of $T$.
\vfill\eject

{\bf References.}

\item {1.} R. Evans, in {\it Les Houches, Session XLVIII, 1988,
            Liquids at Interfaces} (Elsevier, New York, 1989).
\item {2.}  R. Evans, Adv.  Phys. {\bf 28},143 (1979).
\item {3.}  J.M.J.van Leeuwen and J. V. Sengers, Phys. Rev. {\bf A39}, 6346 (1989).
\item {4.}  A. Robledo, J. Stat.Phys. {\bf 89}, 273 (1997). 
\item {5.}  K. R. Mecke and S. Dietrich, Phys. Rev. E {\bf 59},6766 (1999).
\item {6.}  J. Stecki, J. Chem. Phys. {\bf 109}, 5002 (1998); 
           see also ibid. {\bf 114},7574 (2001).
\item {7.}  Jose G. Segovia-Lopez and Victor Romero-Rochin, Phys. Rev. Lett.
            {\bf 86}, 2369 (2001).
\item{8.} M.~Napi\'orkowski and S.~Dietrich, Zeitschrift f. Physik {\bf B97}, 511 (1995) where 
references to earlier work can be found.
           
\item{9.}  R.L.C.Vink, J. Horbach, and K.Binder, J. Chem. Phys. {\bf 122}, 134905 (2005)
\item{10.}  J. Daillant, European Physical Society
 Conference on Liquids, Granada (1999); C. Fradin, A. Braslau, D. Luzet,
 D. Smilgies, M. Alba, M. Boudet, K. Mecke, and J. Daillant,  Nature
{\bf 403}, 871 (2000).
\medskip
\item{11.} H.J.Hilhorst and J.M.J. van Leeuwen, Physica {\bf A107}, 319 (1981). 
\item{12.}  J.Stecki and J. Dudowicz, Proc.Roy.Soc.A (London) {\bf 400},1, 263 (1985).
\item{13.}  J. Dudowicz and J.Stecki, in "Fluid Interfacial phenomena" (C.A.Croxton,Editor)  
               J.Wiley 1986, p. 637ff.
\item {14.} D.B.Abraham, in {\it Phase Transition and Critical Phenomena}\hfill\break
(ed.C.Domb and J.L.Lebowitz),Vol.10, Academic Press,London (1986). 
\item{15.} A.~Ciach, Phys.~Rev.~B {\bf 34}, 1932 (1986). 
\item{16.} A.Ciach, J.Dudowicz, and  J.Stecki, Phys. Rev. Lett. {\bf 56}, 1482 (1986);
\item{17.} A. Kooiman, J.M.J. Van Leeuwen, R.K.P.Zia, Physica {\bf A170}, 124 (1990) 

\vfill\eject
\bigskip
{\bf Figure Captions. }

{\bf Caption to Fig.1} 

A portion of the SOS interface (which must be without overhangs or clusters).
The lattice constant is 1, ordinates are $x =j$ and  
abscissae are $h(x)=h_j \le M$. 

\bigskip
{\bf Caption to Fig.2 }

An example of the projection $\bar H(q)$ plotted against $x\equiv q^2$ 
in a range wide enough to display all three regions; the plateau, the 
main capillary wave region $\bar H \sim 1/x$; and the nearest neighbour
peak at $x \sim 4\pi^2 $. This particular curve for $M=31$ was fitted 
to Eq.(3.1) with $d=0.2964, g=122.40, K_4=-8.0 $. $Q=0.5$ . 

\bigskip
{\bf  Caption to Fig.3 }

 The projections $\bar H(q)$ plotted against $x\equiv q^2$ for the 
 following widths $M$ of the strip M= 120,001; 12,001; 1201; 601; 301;
 121; 61; - from lowest values to the highest in that order. The line
 is $y=1.e-6/x$ to which all $\bar H$ are parallel, perfectly so in 
 the center of the plot. All curves were  fitted to eq.(1.1) and all
 fits produced negative  $K_4$. $Q=0.5$ . 

\bigskip
{\bf  Caption to Fig.4}

The interfacial structure factor $S(x)$ defined by (2.10) plotted 
against  $x\equiv q^2$ for the widths M= 120001(crosses), 
12001(plus signs), 1201(stars), 121(boxes), 
and the line fitted to (1.1). The nearest-neighbor peak and the 
faint increase in $S$ for $q>0.1$ are to be noticed. This faint 
increase forces a negative $\kappa$. $Q=0.1$. The slope is practically 
equal to the stiffness coefficient[17], which explicitely is $(1-Q^2)/(2Q)$.

\vfill\eject
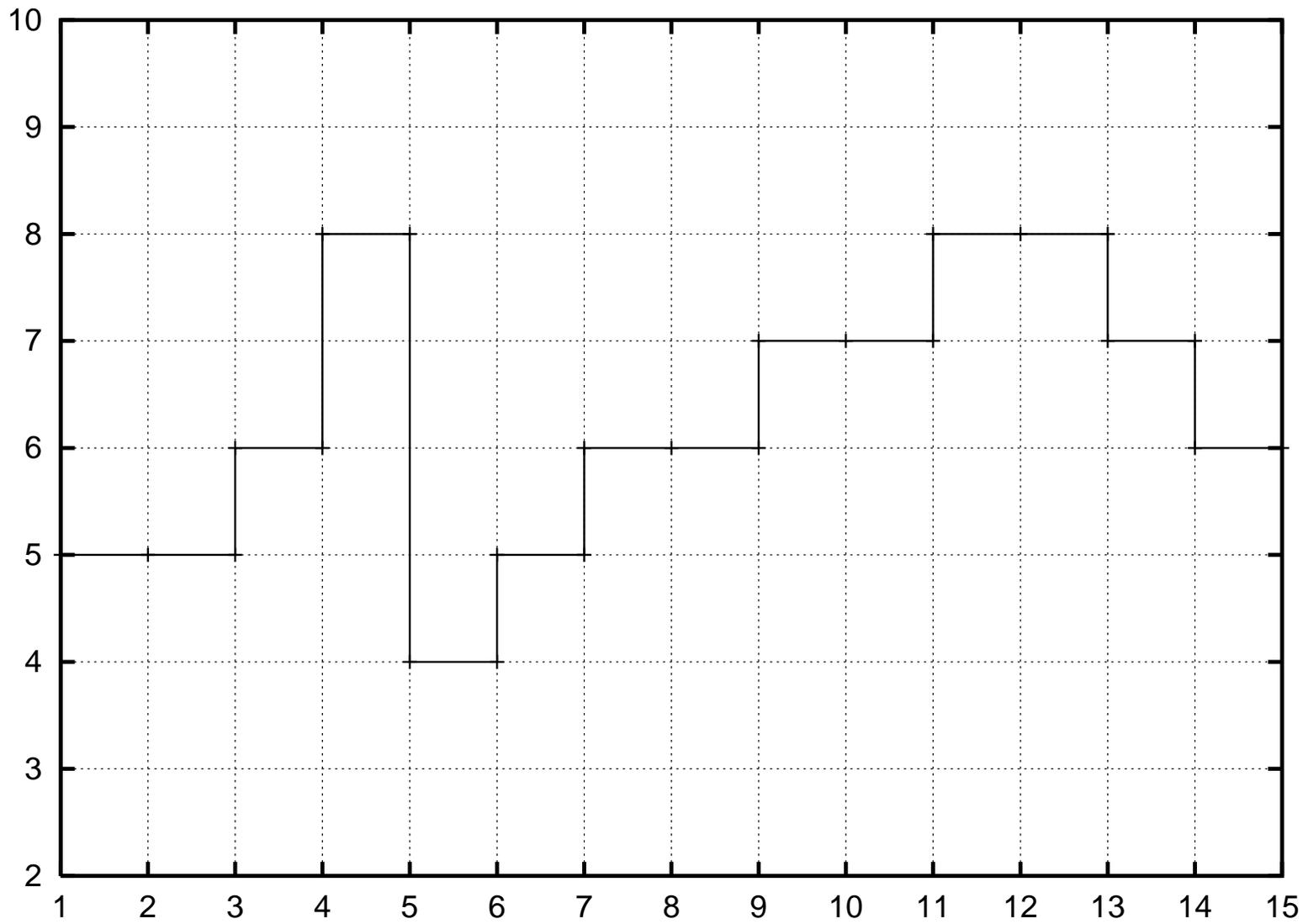
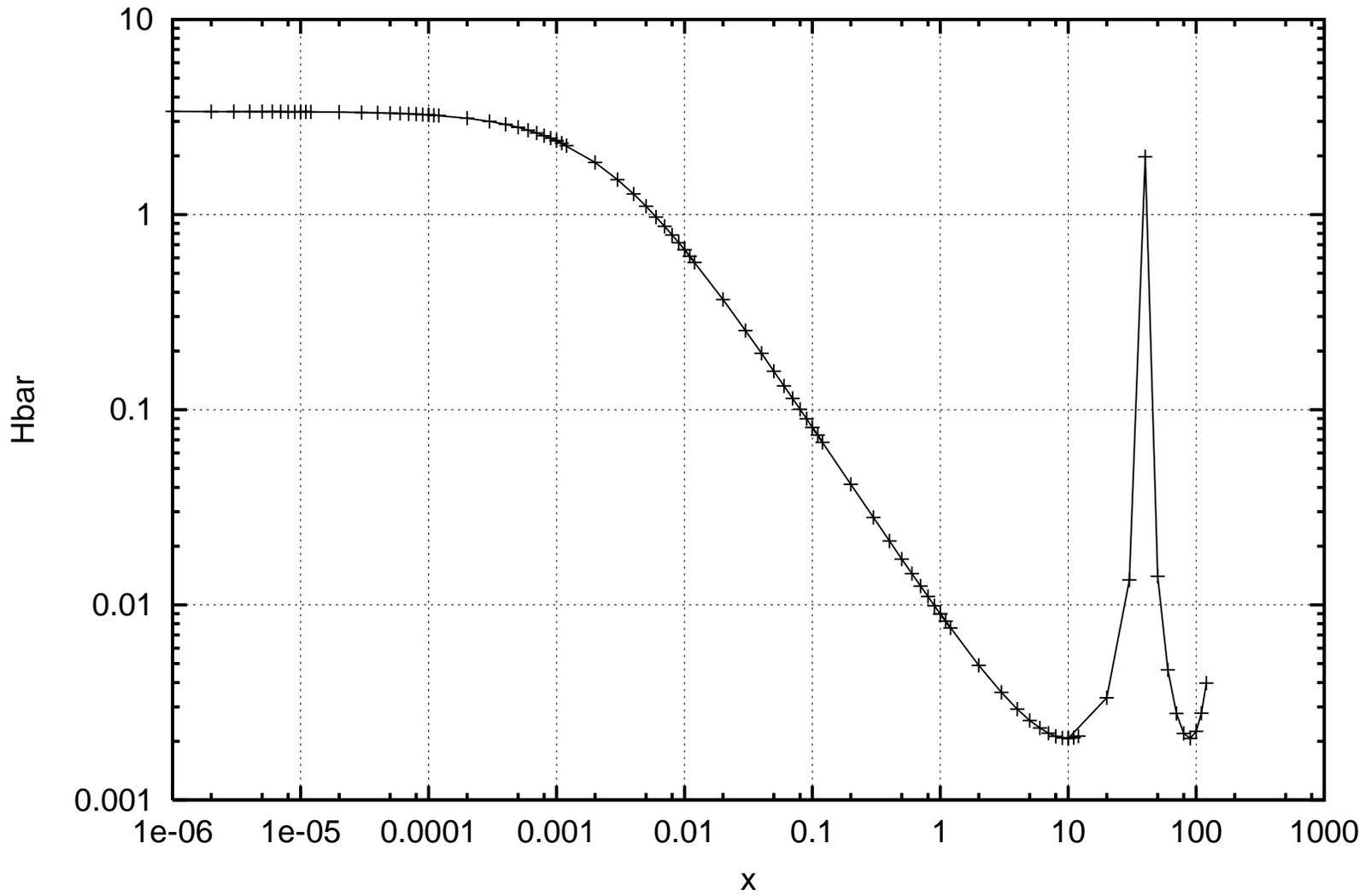
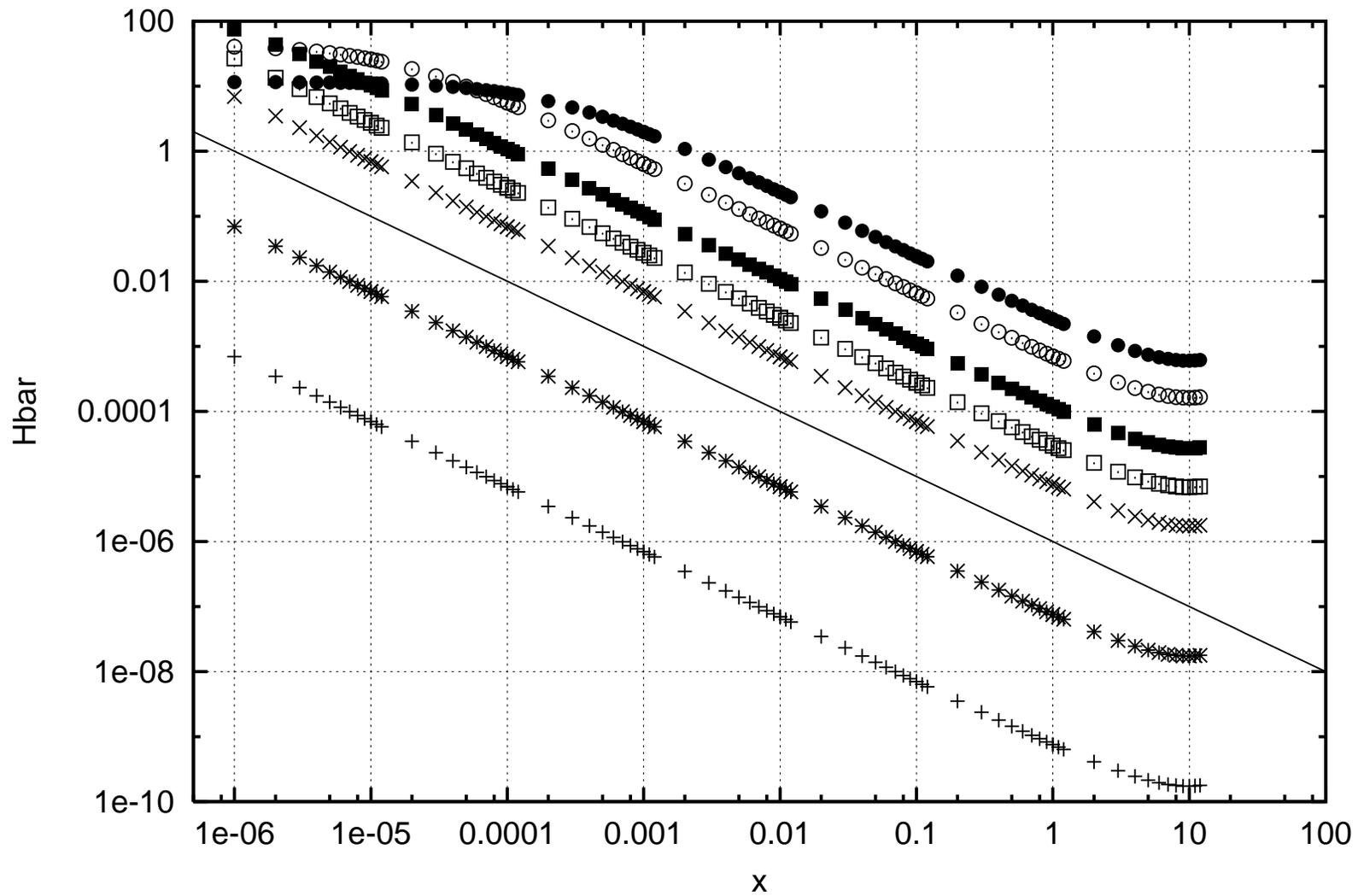
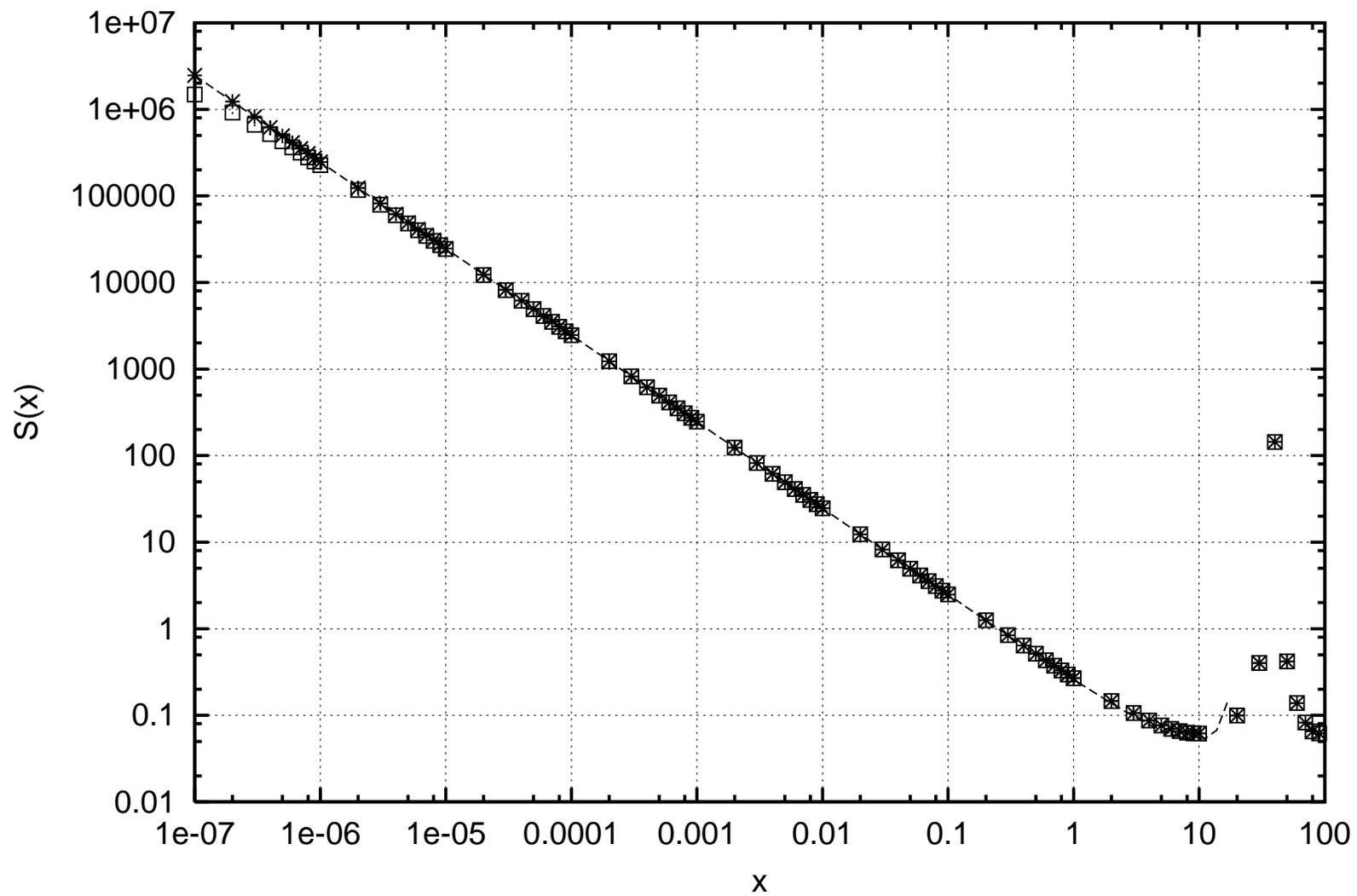
\end